\input harvmac

\newread\epsffilein    
\newif\ifepsffileok    
\newif\ifepsfbbfound   
\newif\ifepsfverbose   
\newif\ifepsfdraft     
\newdimen\epsfxsize    
\newdimen\epsfysize    
\newdimen\epsftsize    
\newdimen\epsfrsize    
\newdimen\epsftmp      
\newdimen\pspoints     
\pspoints=1bp          
\epsfxsize=0pt         
\epsfysize=0pt         
\def\epsfbox#1{\global\def\epsfllx{72}\global\def\epsflly{72}%
   \global\def\epsfurx{540}\global\def\epsfury{720}%
   \def\lbracket{[}\def\testit{#1}\ifx\testit\lbracket
   \let\next=\epsfgetlitbb\else\let\next=\epsfnormal\fi\next{#1}}%
\def\epsfgetlitbb#1#2 #3 #4 #5]#6{\epsfgrab #2 #3 #4 #5 .\\%
   \epsfsetgraph{#6}}%
\def\epsfnormal#1{\epsfgetbb{#1}\epsfsetgraph{#1}}%
\def\epsfgetbb#1{%
%
%
\openin\epsffilein=#1
\ifeof\epsffilein\errmessage{I couldn't open #1, will ignore it}\else
%
%
   {\epsffileoktrue \chardef\other=12
    \def\do##1{\catcode`##1=\other}\dospecials \catcode`\ =10
    \loop
       \read\epsffilein to \epsffileline
       \ifeof\epsffilein\epsffileokfalse\else
%
%
          \expandafter\epsfaux\epsffileline:. \\%
       \fi
   \ifepsffileok\repeat
   \ifepsfbbfound\else
   \ifepsfverbose\message{
     No bounding box comment in #1; using defaults}\fi\fi
   }\closein\epsffilein\fi}%
%
%
%
\def\epsfclipoff{\def\epsfclipstring{\ifepsfdraft\space clip\fi}}%
\epsfclipoff
\def\epsfsetgraph#1{%
   \epsfrsize=\epsfury\pspoints
   \advance\epsfrsize by-\epsflly\pspoints
   \epsftsize=\epsfurx\pspoints
   \advance\epsftsize by-\epsfllx\pspoints
%
%
   \epsfxsize\epsfsize\epsftsize\epsfrsize
   \ifnum\epsfxsize=0 \ifnum\epsfysize=0
      \epsfxsize=\epsftsize \epsfysize=\epsfrsize
      \epsfrsize=0pt
%
%
     \else\epsftmp=\epsftsize \divide\epsftmp\epsfrsize
       \epsfxsize=\epsfysize \multiply\epsfxsize\epsftmp
       \multiply\epsftmp\epsfrsize \advance\epsftsize-\epsftmp
       \epsftmp=\epsfysize
       \loop \advance\epsftsize\epsftsize \divide\epsftmp 2
       \ifnum\epsftmp>0
          \ifnum\epsftsize<\epsfrsize\else
             \advance\epsftsize-\epsfrsize \advance\epsfxsize\epsftmp \fi
       \repeat
       \epsfrsize=0pt
     \fi
   \else \ifnum\epsfysize=0
     \epsftmp=\epsfrsize \divide\epsftmp\epsftsize
     \epsfysize=\epsfxsize \multiply\epsfysize\epsftmp   
     \multiply\epsftmp\epsftsize \advance\epsfrsize-\epsftmp
     \epsftmp=\epsfxsize
     \loop \advance\epsfrsize\epsfrsize \divide\epsftmp 2
     \ifnum\epsftmp>0
        \ifnum\epsfrsize<\epsftsize\else
           \advance\epsfrsize-\epsftsize \advance\epsfysize\epsftmp \fi
     \repeat
     \epsfrsize=0pt
    \else
     \epsfrsize=\epsfysize
    \fi
   \fi
%
%
   \ifepsfverbose\message{#1: width=\the\epsfxsize, height=\the\epsfysize}\fi
   \epsftmp=10\epsfxsize \divide\epsftmp\pspoints
   \vbox to\epsfysize{\vfil\hbox to\epsfxsize{%
      \ifnum\epsfrsize=0\relax
        \includegraphics{\ifepsfdraft}%
      \else
        \epsfrsize=10\epsfysize \divide\epsfrsize\pspoints
        \includegraphics{\ifepsfdraft}%
      \fi
      \hfil}}%
\global\epsfxsize=0pt\global\epsfysize=0pt}%
%
%
{\catcode`\%=12\global\let\epsfpercent=
%
%
\long\def\epsfaux#1#2:#3\\{\ifx#1\epsfpercent
   \def\testit{#2}\ifx\testit\epsfbblit
      \epsfgrab #3 . . . \\%
      \epsffileokfalse
      \global\epsfbbfoundtrue
   \fi\else\ifx#1\par\else\epsffileokfalse\fi\fi}%
%
%
\def\epsfempty{}%
\def\epsfgrab #1 #2 #3 #4 #5\\{%
\global\def\epsfllx{#1}\ifx\epsfllx\epsfempty
      \epsfgrab #2 #3 #4 #5 .\\\else
   \global\def\epsflly{#2}%
   \global\def\epsfurx{#3}\global\def\epsfury{#4}\fi}%
%
%
\def\epsfsize#1#2{\epsfxsize}
%
%

\noblackbox
%
%
%
%
%
\newcount\figno
\figno=0
\def\fig#1#2#3{
\par\begingroup\parindent=0pt\leftskip=1cm\rightskip=1cm\parindent=0pt
\baselineskip=11pt
\global\advance\figno by 1
\midinsert
\epsfxsize=#3
\centerline{\epsfbox{#2}}
\vskip 12pt
\centerline{{\bf Figure \the\figno:} #1}\par
\endinsert\endgroup\par}
\def\figlabel#1{\xdef#1{\the\figno}}
\def\mulfig#1#2#3#4{
\par\begingroup\parindent=0pt\leftskip=1cm\rightskip=1cm\parindent=0pt
\baselineskip=11pt
\global\advance\figno by 1
\midinsert
\centerline{\epsfxsize=#4\epsfbox{#2}\epsfxsize=#4\epsfbox{#3}}
\vskip 12pt
\centerline{{\bf Figure \the\figno:} #1}\par
\endinsert\endgroup\par}
%
%
%
%
%
\def\pano{\par\noindent}

\def\meno{\medskip\noindent}

\font\cmss=cmss10
\font\cmsss=cmss10 at 7pt
\def\rlx{\relax\leavevmode}
\def\inbar{\vrule height1.5ex width.4pt depth0pt}
\def\IC{\relax\,\hbox{$\inbar\kern-.3em{\rm C}$}}
\def\IN{\relax{\rm I\kern-.18em N}}
\def\IP{\relax{\rm I\kern-.18em P}}
\def\ZZ{\rlx\leavevmode\ifmmode\mathchoice{\hbox{\cmss Z\kern-.4em Z}}
 {\hbox{\cmss Z\kern-.4em Z}}{\lower.9pt\hbox{\cmsss Z\kern-.36em Z}}
 {\lower1.2pt\hbox{\cmsss Z\kern-.36em Z}}\else{\cmss Z\kern-.4em Z}\fi}
\def\narrowplus{\kern -.04truein + \kern -.03truein}
\def\narrowminus{- \kern -.04truein}
\def\narrowminussub{\kern -.02truein - \kern -.01truein}
\def\a{\alpha}
\def\b{\beta}
\def\si{\sigma}
\def\cl{\centerline}

\def\o#1{\overline{#1}}
\def\v#1{\vec{#1}}
\def\pt{\partial}
\def\lra{\longrightarrow}

\def\lb{\{ }
\def\rb{\} }
\def\la{\langle}
\def\ra{\rangle}

\def\qi{\vec{q}_i}
\def\qa{\vec{q}_a}
\def\z{\vec{z}}

\def\lg{Landau-Ginzburg }

\def\n{\nu}
\def\sqr#1#2{{\vcenter{\vbox{\hrule height.#2pt
            \hbox{\vrule width.#2pt height#1pt \kern#1pt
                  \vrule width.#2pt}\hrule height.#2pt}}}}
\def\square
{\mathop{\mathchoice{\sqr{12}{15}}{\sqr{9}{12}}{\sqr{6.3}{9}}{\sqr{4.5}{9}}}}
%
%
%
%
%
\lref\rbw{R. Blumenhagen and A. Wi{\ss}kirchen, Nucl. Phys. {\bf B454} 
  (1995) 561.}
\lref\rbwmod{R. Blumenhagen and A. Wi{\ss}kirchen, Nucl. Phys. {\bf B475}
  (1996) 225.}
\lref\rbs{R. Blumenhagen and S. Sethi, hep--th/9611172.}
\lref\rduality{J. Distler and S. Kachru, Nucl. Phys. {\bf B442} (1995) 64.}
\lref\rdkmodul{J. Distler and S. Kachru, Nucl. Phys. {\bf B430} (1994) 13.}
\lref\rbswmirror{R. Blumenhagen, R. Schimmrigk and A. Wi{\ss}kirchen,
  hep--th/9609167.}
\lref\rbsw{R. Blumenhagen, R. Schimmrigk and A. Wi{\ss}kirchen,
  Nucl. Phys. {\bf B461} (1996) 460.}
\lref\rgp{B. Greene and R. Plesser, Nucl. Phys. {\bf B338} (1990) 15.}
\lref\rew{E. Witten, Nucl. Phys. {\bf B403} (1993) 159.}
\lref\rkm{T. Kawai and K. Mohri, Nucl. Phys. {\bf B425} (1994) 191.}
\lref\rdgm{J. Distler, B. Greene and D. Morrison, 
  Nucl.Phys. {\bf B481} (1996) 289 \semi
  Ti-Ming Chiang, J. Distler and B. Greene, hep--th/9602030.}
\lref\rdnotes{J. Distler, ``Notes on $(0,2)$ Superconformal Field Theories'',
  published in Trieste HEP Cosmology (1994) 322.}
\lref\rdk{J. Distler and S. Kachru, Nucl. Phys. {\bf B413} (1994) 213.}
\lref\rwita{E. Witten, Nucl. Phys. {\bf B268} (1986) 79.}
\lref\rdsww{M. Dine, N. Seiberg, X. Wen and E. Witten, 
  Nucl. Phys. {\bf B278} (1986) 769; ibid. {\bf B289} (1987) 319.}
\lref\rdga{J. Distler and B. Greene, Nucl. Phys. {\bf B304} (1988) 1.}
\lref\rdgb{J. Distler and B. Greene, Nucl. Phys. {\bf B309} (1988) 295.}
\lref\rkw{S. Kachru and E. Witten, Nucl. Phys. {\bf B407} (1993) 637.}
\lref\rkt{M. Kreuzer and M. Nikbakht-Tehrani, hep--th/9611130.} 
\lref\rnt{M. Nikbakht-Tehrani, hep--th/9612067.}
\lref\rdix{L. Dixon, Lectures given at the 1987 ICTP Summer Workshop
  in High Energy Physics and Cosmology.}
\lref\rcand{P. Candelas, X. de la Ossa, P. Green and L. Parks, 
  Nucl. Phys. {\bf B359} (1991) 21.}  
\lref\rvlg{C. Vafa, Mod. Phys. Lett. {\bf A4} (1989) 1169.}
\lref\rdiscrete{C. Vafa, Nucl. Phys. {\bf B273} (1986) 592.}
\lref\rwlg{E. Witten, Int. J. Mod. Phys. {\bf A9} (1994) 4783.}
\lref\rFT{C Vafa, Nucl. Phys. {\bf B469} (1996) 403 \semi
  D.R. Morrison and C. Vafa, Nucl. Phys. {\bf B473} (1996) 74, 
  Nucl. Phys. {\bf B476} (1996) 437 \semi 
  S. Sethi, E. Witten, C. Vafa, Nucl. Phys. {\bf B480} (1996) 213 \semi
  I. Brunner, M. Lyncker and R. Schimmrigk, Phys. Lett. {\bf B387} (1996) 
  750 \semi
  R. Friedman, J. Morgan and E. Witten,  hep--th/9701162\semi
  M. Bershadsky, A. Johansen, T. Pantev and  V. Sadov, hep--th/9701165.}
\lref\rvi{K. Intriligator and C. Vafa, Nucl. Phys. {\bf B339} (1990) 95.}
\lref\rsw{E. Silverstein and E. Witten, Nucl. Phys. {\bf B444} (1995) 161.}
\lref\rks{M. Kreuzer and H. Skarke, Mod. Phys. Lett. {\bf A10} (1995) 1073.}
\lref\rdhvw{L. Dixon, J. Harvey, C. Vafa, and E. Witten, Nucl. Phys. 
  {\bf B274} (1986) 285.}
\lref\rkyy{ T. Kawai, Y. Yamada, and S-K Yang, Nucl. Phys. {\bf B414} 
  (1994) 191.}
\lref\rvw{C. Vafa and E. Witten, J. Geom. Phys. {\bf 15} (1995) 189.}
\lref\rbconjecture{V. Batyrev, J. Alg. Geom. {\bf 3} (1994) 493.}
\lref\rsyz{A. Strominger, S-T Yau, and E. Zaslow, hep--th/9606040.}
\lref\rbh{P. Berglund and M. Henningson, Nucl. Phys. {\bf B433} (1995) 311.}
\lref\rwitF{E. Witten,  Nucl. Phys. {\bf B474} (1996) 343.}
\lref\rper{P. Berglund, C.V. Johnson, S. Kachru, P. Zaugg, Nucl. Phys.
  {\bf 460} (1996) 252.}
%
%
%
%
%
\Title{\vbox{\hbox{hep-th/9702199}
             \hbox{IASSNS--HEP--97/13}}}
{Aspects of $(0,2)$ Orbifolds and Mirror Symmetry}
\smallskip
\centerline{{Ralph Blumenhagen${}^1$}  and  {Michael Flohr${}^2$} }
\bigskip
\centerline{${}^{1,2}$ \it School of Natural Sciences,
                       Institute for Advanced Study,}
\centerline{\it Olden Lane, Princeton NJ 08540, USA}
\smallskip
\bigskip
\bigskip\bigskip
\centerline{\bf Abstract}
\noindent
We study orbifolds of $(0,2)$ models and their relation to $(0,2)$
mirror symmetry. In the Landau-Ginzburg phase of a $(0,2)$ model the
superpotential features a whole bunch of discrete symmetries, which
by quotient action lead to a variety of consistent $(0,2)$ vacua.
We study a few  examples in very much detail. Furthermore, we comment
on the application of $(0,2)$ mirror symmetry to the calculation
of Yukawa couplings in the space-time superpotential.
\phantom{\rwita\rdsww\rdga\rew\rkw\rdk\rdkmodul\rduality\rdnotes\rkm\rsw\rbw
\rper\rbsw\rbwmod\rdgm\rkt\rnt}
\footnote{}
{\pano
${}^1$ e--mail:\ blumenha@sns.ias.edu, ~~supported by NSF grant 
PHY--9513835
\pano
${}^2$ e--mail:\ flohr@sns.ias.edu, ~~supported by the 
Deutsche Forschungsgemeinschaft
\pano}
\Date{02/97}
%
%
%
%
%
%
\newsec{Introduction}

Despite gradual progress in revealing  the existence and structure of 
phenomenologically promising $(0,2)$
world-sheet supersymmetric compactifications of the heterotic string [1-18], 
the knowledge we have is still far less compared to their more prominent 
left-right symmetric subset of $(2,2)$ models. Since some special 
elliptically fibered $(0,2)$ 
models in both six and four dimensions made their appearance in conjectured 
F-theory, heterotic string dualities \rFT, to have a better understanding, 
in particular of their moduli spaces, clearly is desirable. In the $(2,2)$ 
case, a combination of some non-renormalization theorems for certain 
couplings in the superpotential \rdix,\rdgb\  and mirror symmetry served as 
powerful tools for formulating an exact geometric  description of the complex
and K\"ahler moduli spaces \rcand .
\pano
In the $(0,2)$ case we are on much looser ground. On the one hand, the proof 
of exactness of certain Yukawa couplings in the large radius limit heavily
relied on the left-moving world sheet $N=2$ supersymmetry. Furthermore, for
small radius there does not even exist an algebraic distinction among the
possible complex, K\"ahler and bundle moduli. On the other hand, $(0,2)$
mirror symmetry is still in its infancy. Even though for a special subset
of $(0,2)$ models strong indications of mirror symmetry have been found
in \rbswmirror, we do not know whether this duality extends to more general 
$(0,2)$ compactifications. 
\pano
With this background in mind, in this letter we investigate further the 
implementation of mirror symmetry in the $(0,2)$ context and its application 
to the calculations of certain 3-point couplings. Recently, a description of 
orbifolds of $(0,2)$ models in their Landau-Ginzburg phase has been presented
\rbs. In particular, a formula for the elliptic genus
in the quotient model has been derived. In contrast to the $(2,2)$ case,
a priori the $(0,2)$ superpotential has an infinite number of discrete
symmetries subject to some anomaly constraints. After reviewing  the basics
of the orbifold construction, we systematically study  $(0,2)$  orbifolds of
a $(0,2)$ orbifold descendant of the $(2,2)$ quintic in very much detail.
We find that even by modding out in each case  only
one discrete symmetry one ends up with a large number of different models
showing (almost) mirror symmetry. As a by-product we find that the
simple current construction given in \rbw\ is nothing else than a $Z_2$
$(0,2)$ orbifold of a $(2,2)$ model. This gives a way of constructing 
consistent $(0,2)$ models as orbifold descendants of $(2,2)$ models. 
In order to see whether $(0,2)$ mirror symmetry is only an artifact for 
such descendants of $(2,2)$ models, we also study an example which 
is not supposed to be of this type.
\pano
In the last section we draw the  minimal conclusion from mirror symmetry,
allowing us to derive simple selection rules for  some Yukawa couplings of 
the form $\la 10,\o{16},\o{16}\ra$ in the case of $SO(10)$ gauge group.  

\newsec{Review of $(0,2)$ Orbifolds}

We consider $(0,2)$ models described by linear $\si-$ models \rew\ which for
small radius $r\ll 0$ are equivalent  to $(0,2)$ Landau-Ginzburg models
\footnote{$^1$}{The restriction to linear $\si-$ models might exclude
a lot of the elliptically fibered models naturally arrising in recent
F-theory/heterotic dualities. As shown in \rsw\ the former models
are generically
not subject to world-sheet instanton corrections of the space time
superpotential, whereas such a feature is expected from special divisors
in F-theory \rwitF }.  
There is a number of chiral superfields:  $\lb\Phi_i\vert i=1,\ldots,N\rb$ 
and a number of Fermi superfields: 
$\lb\Lambda^a\vert a=1,\ldots,M=N_a+N_j\rb$ 
which are governed by a superpotential of the form
\eqn\superpot{ W = \Lambda^a F_a(\Phi_i)+\Lambda^{N_a+j} W_j(\Phi_i)\,, }
where $W_j$ and $F_a^l$ are quasi-homogeneous polynomials.
In the large radius limit the $W_j$ define hypersurfaces in a weighted 
projective space and the $F_a$ define a vector bundle on this space. For 
appropriate choices of the constraints $W_j$ and $F_a$, the 
superpotential has an isolated singularity at the origin and is 
quasi-homogeneous of degree one,
if one assigns charges $\omega_i/m$ to $\Phi_i$,
$n_a/m$ to $\Lambda_a$, and $1-d_j/m$ to $\Lambda_{N_a+j}$.
Quasi-homogeneity implies the existence of a right-moving $R$-symmetry, and a
left-moving  $U(1)_L$. The associated currents are denoted by $J_R$ and 
$J_L$, respectively.
The charges of the various fields with respect to
these $U(1)$ currents are summarized in the following table:
\vskip 0.1in
\meno
\cl{\vbox{
\hbox{\vbox{\offinterlineskip
\def\tablespace{height2pt&\omit&&\omit&&\omit&&\omit&&\omit&\cr}
\def\tablerule{\tablespace\noalign{\hrule}\tablespace}

\hrule\halign{&\vrule#&\strut\hskip0.2cm\hfil#\hfill\hskip0.2cm\cr
\tablespace
& Field && $\phi_{i}$ && $\psi_{i}$ && $\lambda_a$  && $\lambda_{N_a+j}$ &\cr
\tablerule
& $q_L$ && ${\omega_i\over m}$ &&${\omega_i\over m}$  && ${n_a \over m}-1$
&& $-{d_j\over m}$  &\cr
\tablerule
& $q_R$ && ${\omega_i\over m}$ && ${\omega_i\over m}-1$ && ${n_a \over m}$
&& $1-{d_j \over m}$ &\cr
\tablespace}\hrule}}}}
\cl{
\hbox{{\bf Table 1:}{\it ~~Left and right charges of the
fields in the LG model.}}}
\meno
Of course, the fermions, $\psi_i$, belong to the chiral superfield, $\Phi_i$,
while the fermions, $\lambda_a$  are the lowest components of
the Fermi superfields $\Lambda_a$.
Anomaly cancellation for  these two global $U(1)$ symmetries is equivalent to 
the anomaly conditions expected from the large radius limit:
\eqn\anfree{\eqalign{ &\sum \omega_i = \sum d_j\,,\quad\quad 
                       \sum n_a = m\,,\cr
                      &\sum d_j^2 - \sum w_i^2 = m - \sum n_a^2\,.\cr
}}
{}For appropriate choices of the functions $W_j$ and $F_a$, in general there
exist a bunch of discrete symmetries of the superpotential acting on the
fields as
\eqn\phase{ \Phi_i\rightarrow e^{2\pi i q_i} \Phi_i,
            \Lambda_a\rightarrow e^{ -2\pi i q_a} \Lambda_a\,.}
This defines a $\ZZ_h$ action on the fields, where $h$ is the minimal
common denominator of the the charges $q_i$, $q_a$. In general one has
multiple quotient actions of order $h^0,\ldots,h^{P-1}$, where the first 
quotient should be
the GSO projection $\ZZ_m$. Then one can define the following quantities:
\eqn\Rmatrix{ R^{\mu\nu}=\sum_{a=1}^M q^{\mu}_a q^{\nu}_a -
                        \sum_{i=1}^N q^{\mu}_i q^{\nu}_i\,,\quad\quad
                 r^\mu =\sum_{a=1}^M q^\mu_a - \sum_{i=1}^N q^\mu_i\,. }
As was shown in \rbs\ the orbifold partition function 
can be written as a sum over all twisted sectors as
\eqn\orbieg{ Z_{orb}(\tau,\n)={1\over \prod{ h^\mu}}
       \sum_{\alpha^0,\beta^0=0}^{h^0-1}
       \ldots \sum_{\alpha^{P-1},\beta^{P-1}=0}^{h^{P-1}-1}\, 
       \epsilon(\v{\a},\v{\b})\,\,
       {\scriptstyle{\vec\beta\,}}{\square_{\vec\alpha}}\,(\tau,\n,0)\,,}
and is modular invariant only if the phase factor
\eqn\dismodb{ \epsilon(\vec\alpha,\vec\beta)=e^{\pi i \vec{w}(\vec\alpha +
              \vec\beta)}\, e^{\pi i \vec\alpha Q \vec\beta } }
satisfies
\eqn\dismodc{\eqalign{
  &Q^{\mu\n} + Q^{\n\mu} \in 2\ZZ\,, \quad\quad
   w^\mu + Q^{\mu\mu} \in 2\ZZ\,, \cr
  &(w^\mu - r^\mu)h^\mu = 0 \quad {\rm mod}\ 2\,, \quad\quad
   (Q^{\mu\n} + R^{\mu\n}) h^\n = 0 \quad {\rm mod}\ 2\,, \cr
}}
{}for any $\mu,\nu\in\{0,\ldots,P-1\}$.
These conditions provide
constraints on $r^\mu$ and $R^{\mu\mu}$:
\eqn\even{ \eqalign{ 
  r^\mu h^\mu \in \cases{2\ZZ\ {\rm for}\ h^\mu\ {\rm even\,,} \cr
                          \ZZ\ {\rm for}\ h^\mu\ {\rm odd\,,} \cr} \quad\quad
  R^{\mu\mu} h^\mu \in\cases{2\ZZ\ {\rm for}\ h^\mu\ {\rm even\,,} \cr
                              \ZZ\ {\rm for}\ h^\mu\ {\rm odd\,.} \cr} 
}}
{}For the off-diagonal terms one obtains the condition
\eqn\offd{ R^{\mu\nu} ={\ZZ \over h^\mu}+{\ZZ \over h^\nu}\,. }
In order for the quotient theory to be used as the internal sector of
a heterotic string theory, there are some further conditions on the charges
that have to be satisfied. 
The difference of the left and right moving $U(1)$ charges in every single
twisted sector must be an integer:
\eqn\conda{ \sum_a \qa - \sum_i \qi \in \ZZ^P\,.}
The gauginos form the untwisted sector must not be projected out
\eqn\condb{ \vec{w}=\left( \sum_a \qa - \sum_i \qi \right)\quad {\rm 
mod}\ 2\,.}
Lastly, we want our canonical projection onto states with left-moving charge
$q_L={1\over 2} r^0 \, {\rm mod}\ \ZZ$. This requirement leads to the
condition,
\eqn\condc{    (Q^{\mu 0}-R^{\mu 0}) \in 2\ZZ\,. }
This determines $Q^{\mu 0}$ in terms of $R^{\mu 0}$ mod 2.
\pano
The massless sector of the orbifold contributes only to the so-called 
$\chi_y$ genus defined as 
\eqn\chiy{ \chi_y = \lim_{q\rightarrow 0} \, (i)^{N-M} q^{ N-M\over 12}
y^{{1\over 2} r^0} Z_{orb}(q,y)\,. }
We denote the contribution to $\chi_y$ from a twisted sector $\v{\a}$ by
$\chi^{\v{\a}}_y$. The contribution from each twisted sector is determined 
in terms of the function,
\eqn\chiyorb{ f^{\v\alpha}(\vec{z}) =(-1)^{\vec{w}\vec\alpha} 
      e^{2\pi i\z\vec{Q}_{\vec\alpha}}
      q^{E_{\vec\alpha}}
      {\prod_a (-1)^{[\vec\alpha\qa]}
      (1 - e^{2\pi i \z\qa} q^{\lb\vec\alpha\qa\rb} )
      (1 - e^{-2\pi i \z\qa} q^{1-\lb\vec\alpha\qa\rb} ) \over
      \prod_i (-1)^{[\vec\alpha\qi]}
      (1- e^{2\pi i \z\qi} q^{\lb\vec\alpha\qi\rb} )
      (1- e^{-2\pi i \z\qi} q^{1-\lb\vec\alpha\qi\rb} ) }\,, }
where $\chi^{\v{\a}}_y$ is given by expanding $f^{\v{\a}} (\v{z})$ in powers 
of $q$, and retaining terms of the form 
$q^0 e^{-2\pi i \z(\vec\sigma+\vec{n})}$,
where $\v{n}\in\ZZ^P$ and $\v\sigma = {1\over 2} \v{w} + {1\over 2} \v{\a} 
(Q - R)$.
Finally, we set $z_1=\ldots=z_{P-1}=0$. Furthermore, we have used the 
abbreviation $\{x\}=x-[x]$ in \chiyorb. The fractionalized
charges and energies in the twisted sectors are given by the formulae:
\eqn\charges{ \eqalign{ 
  \v{Q}_{\vec\alpha}&=
    \sum_a \v{q}_a (\vec\alpha\qa-[\vec\alpha\qa]-{1\over 2}) - 
    \sum_i \v{q}_i (\vec\alpha\qi-[\vec\alpha\qi]-{1\over 2})\,, \cr
  E_{\vec\alpha}&= 
    {1\over 2}\sum_a 
    (\vec\alpha\qa-[\vec\alpha\qa]-1) (\vec\alpha\qa-[\vec\alpha\qa]) - 
    {1\over 2}\sum_i 
    (\vec\alpha\qi-[\vec\alpha\qi]-1) (\vec\alpha\qi-[\vec\alpha\qi])\,. \cr
}}
This is a formula which can easily be put onto a computer, making more 
excessive calculations feasible. 

\newsec{Some special quotients}

\subsec{ $(0,2)$ quotients of $(2,2)$ models}

In this section we apply the orbifolding procedure to some special
models, leading to interesting aspects of $(0,2)$ models. 
First, we consider $(2,2)$ models which are given by a hypersurface
in a weighted projective space $\IP_{\omega_1,\ldots,\omega_5}[d]$. 
Let us transform  such a model to a $(0,2)$ model with data
\eqn\ttmodel{ V(\omega_1,\ldots,\omega_5;d)\lra 
              \IP_{\omega_1,\ldots,\omega_5}[d]\,. } 
Note that in the \lg phase the superpotential is
\eqn\superlg{ W=\sum_{i=1}^5 \Lambda_i {\pt P\over \pt \phi_i} +
                             \Lambda_6\, P \,.}
with $P$ being a transversal polynomial of degree $d$.
Calculating the massless spectrum of such a model gives exactly the
$(2,2)$ result with  extra gauginos occurring in a twisted and the untwisted
sector extending the gauge group from $SO(10)$ to $E_6$. By decoupling
the left moving fermions $\lambda$ from the bosons $\phi$ we are free
to consider also general $(0,2)$ orbifolds. If $d/\omega_1=2l+1$ is odd
we deform the superpotential \superlg\ to
\eqn\superlgb{ W=\Lambda_1 \phi_1^{2l} +  \sum_{i=2}^5 \Lambda_i 
                  {\pt P\over \pt \phi_i} +  \Lambda_6\, \phi_1^{2l+1} }
and divide by the following $\ZZ_2$ action
\eqn\move{ J=\left({2l+1\over 2},0,0,0,0;0,0,0,0,0,{2l-3\over 2} \right) }
which satisfies all anomaly conditions \even \offd. By calculating a
few examples one finds that this orbifold corresponds exactly to the
implementation of the simple current
\eqn\simple{ (\v{q}_i,\v{q}_a)=(0\ 2l+1\ 1)(0\ 0\ 0)^4 (1)(0) }
into the conformal field theory partition function introduced in \rbw.
Thus, following
\rbsw, the move from a $(2,2)$ model $\IP_{\omega_1,\ldots,\omega_5}[d]$ 
to a $(0,2)$ model 
\eqn\movemod{ V(\omega_1,\ldots,\omega_5;d) \lra  
\IP_{2\omega_1,l\omega_1,\omega_2,\ldots,\omega_5}[(l+2)\omega_1,2l\omega_1]}
can be described as a $(0,2)$ orbifold of the $(2,2)$ model. 
Analogously, one expects that $(2,2)$ models can produce different kinds
of $(0,2)$ models via quotient actions. Thus, orbifolding  provides a nice 
way of constructing $(0,2)$ descendants out of $(2,2)$ models.  

\subsec{ Classifying all $(0,2)$ orbifolds of the quintic}

Exactly for the class of models reinterpreted as orbifolds in the last
section, mirror symmetry has been investigated in \rbsw. If the
$(2,2)$ model is of Fermat type, it has been shown by Greene and Plesser 
\rgp\ that
orbifolding by the maximal discrete symmetry (preserving the left moving
$N=2$ symmetry) leads to the mirror model. Even more striking, 
successive orbifolding
leads to a completely mirror symmetric set of vacua. 
We want to see whether a similar pattern also holds in the $(0,2)$
context. In \rbs\ some orbifolds
of the $(0,2)$ descendant of the quintic
\eqn\modella{ V(1,1,1,1,1;5)\lra \IP_{1,1,1,1,2,2}[4,4] }
have been constructed. Successive modding by a few  generating
$\ZZ_5$ orbifolds and introducing non-trivial discrete torsion has lead
to a mirror symmetric subset of all $(0,2)$ orbifold models. In this section
we want to be  more ambitious and start a classification of all possible
discrete symmetries of the superpotential
\eqn\superquint{ W=\sum_{i=1}^4 \lambda_i \phi_i^4 + \lambda_5\phi_5^2
                 + \lambda_6\phi_6^2+ \lambda_7\phi_5\phi_6\,. }
As a first observation, the decoupling of fermionic and bosonic degrees of
freedom in the superpotential allows for an infinite set of solutions of
the Diophantine equations encoding the conditions for its invariance and
the consistency of the model. In
particular, for almost any $n$ we have a non-empty set of solutions with
$\ZZ_n$ symmetry. However, the set of different models with different spectra
is finite, but much larger as in the case of $(2,2)$ models. 
Due to the torsion $Q^{\mu,0}$ determined by the condition  \condc, 
one can restrict the search for solutions with $\ZZ_n$
symmetry to integers modulo $n$ such that for each $n$ there are only 
finitely many possibilities to check. 
\pano
As a second observation, we note that e.g.\ all the orbifold models of the
$(0,2)$ descendant of the quintic
constructed by successive modding by certain $\ZZ_5$ symmetries can also be
found by modding just once with a higher symmetry, for example $\ZZ_{15}$ or 
$\ZZ_{25}$. Led by the theory of induced representations
{}for poly-cyclic groups,
we conjecture that {\it all\/} $(0,2)$ orbifolds to a given basis model can
be obtained by modding out just one (suitable high) symmetry $\ZZ_n$.
\pano
{}For the $(0,2)$ descendant of the quintic we found all possible orbifold
solutions with one $\ZZ_n$ symmetry modded out, $2\leq n\leq 34$. This yields
$71940$ models, but only $179$ different $SO(10)$ spectra and $82$ different
$E_6$ spectra as well as one $SO(12)$ orbifold model with $N_{32} = 
N_{\o{32}} = 6$. There are non-trivial
solutions for all these $n$ except $n=2,4$. The set of different orbifold
models obtained so far is certainly not yet complete, but it is almost.
Assuming the correctness of our conjecture above, 
and keeping in mind that the naive symmetry of the quintic is
$\ZZ_5$, we conjecture that $\ZZ_n$ orbifolds with an upper limit $n=125=5^3$
would exhaust the complete set of orbifold models -- which, however, is 
outside our computation abilities.
\pano
The main observation now is that already our yet incomplete set does
enjoy mirror symmetry to a surprisingly high extent, if $SO(10)$ models
are considered. The situation for $E_6$ models is much less clear, which
mainly is due to the limited ability to read off gauginos in untwisted
sectors using the method introduced in \rbswmirror.
Figures 1 and 2 present our results.

\fig{{\it ~~The almost complete set of orbifolds for the $(0,2)$ descendant 
of the quintic.}}{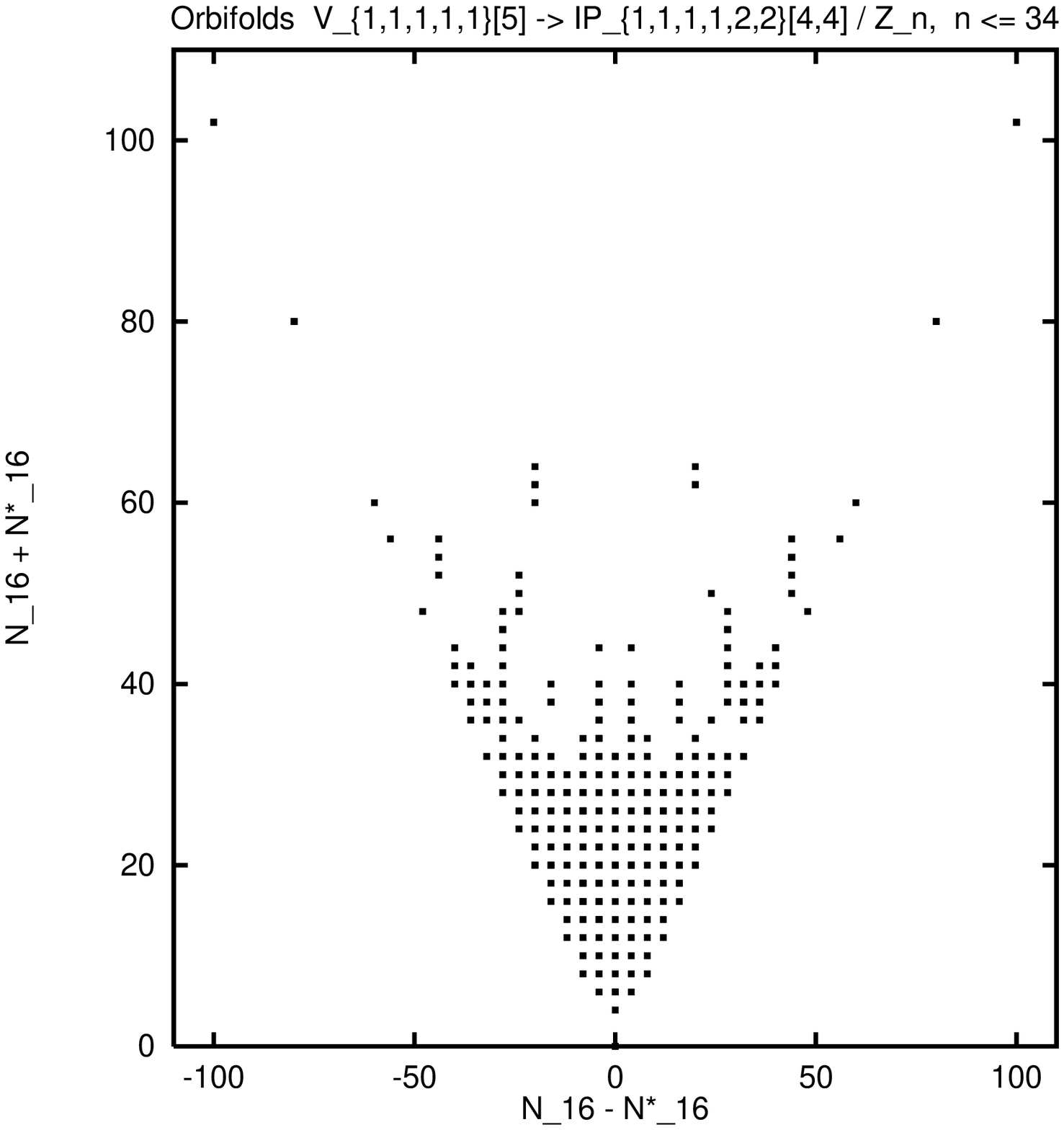}{9.6truecm}
\mulfig{{\it ~~Orbifold models of the $(0,2)$ quintic split into $SO(10)$ 
and $E_6$ gauge groups.}}{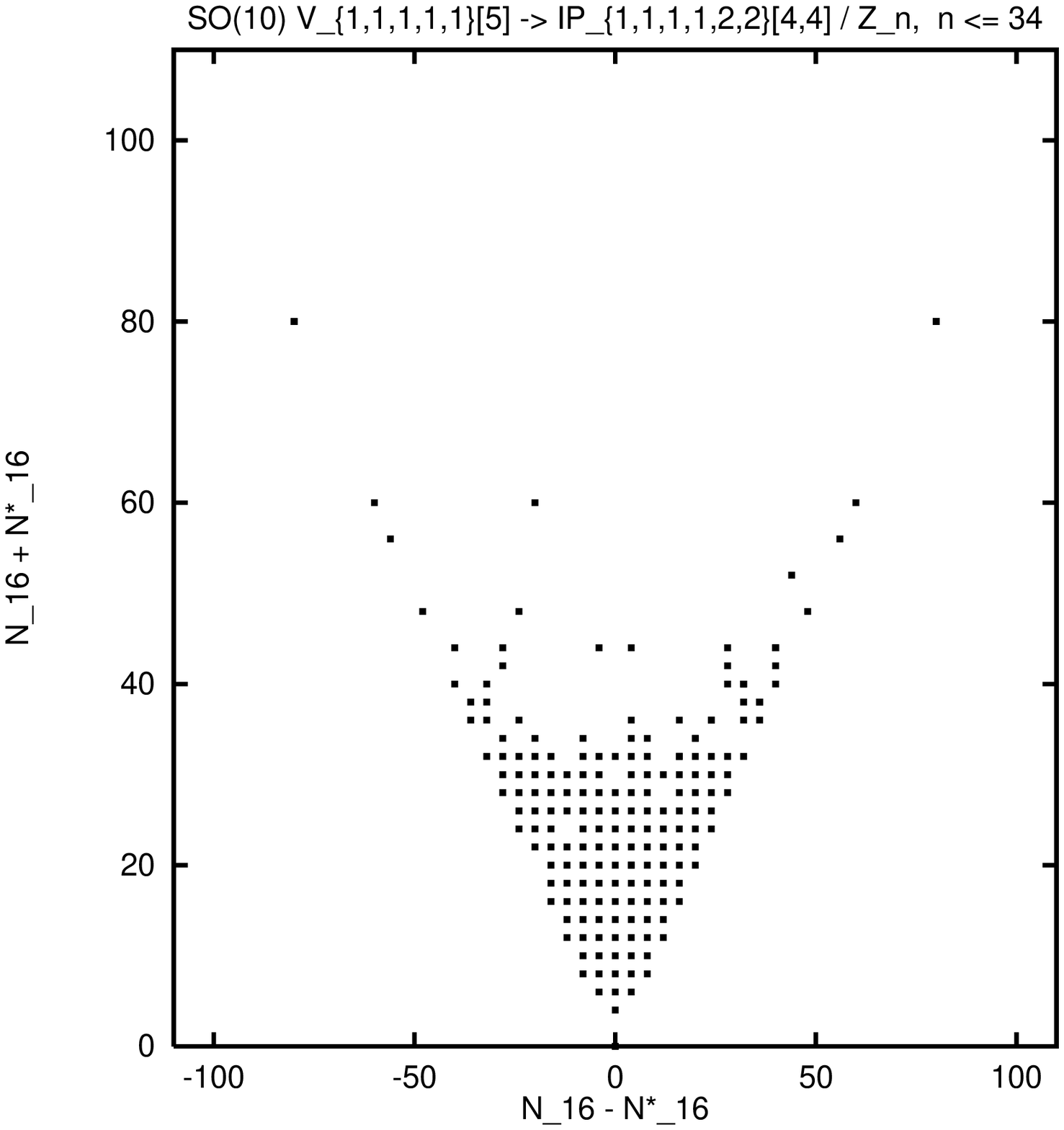}{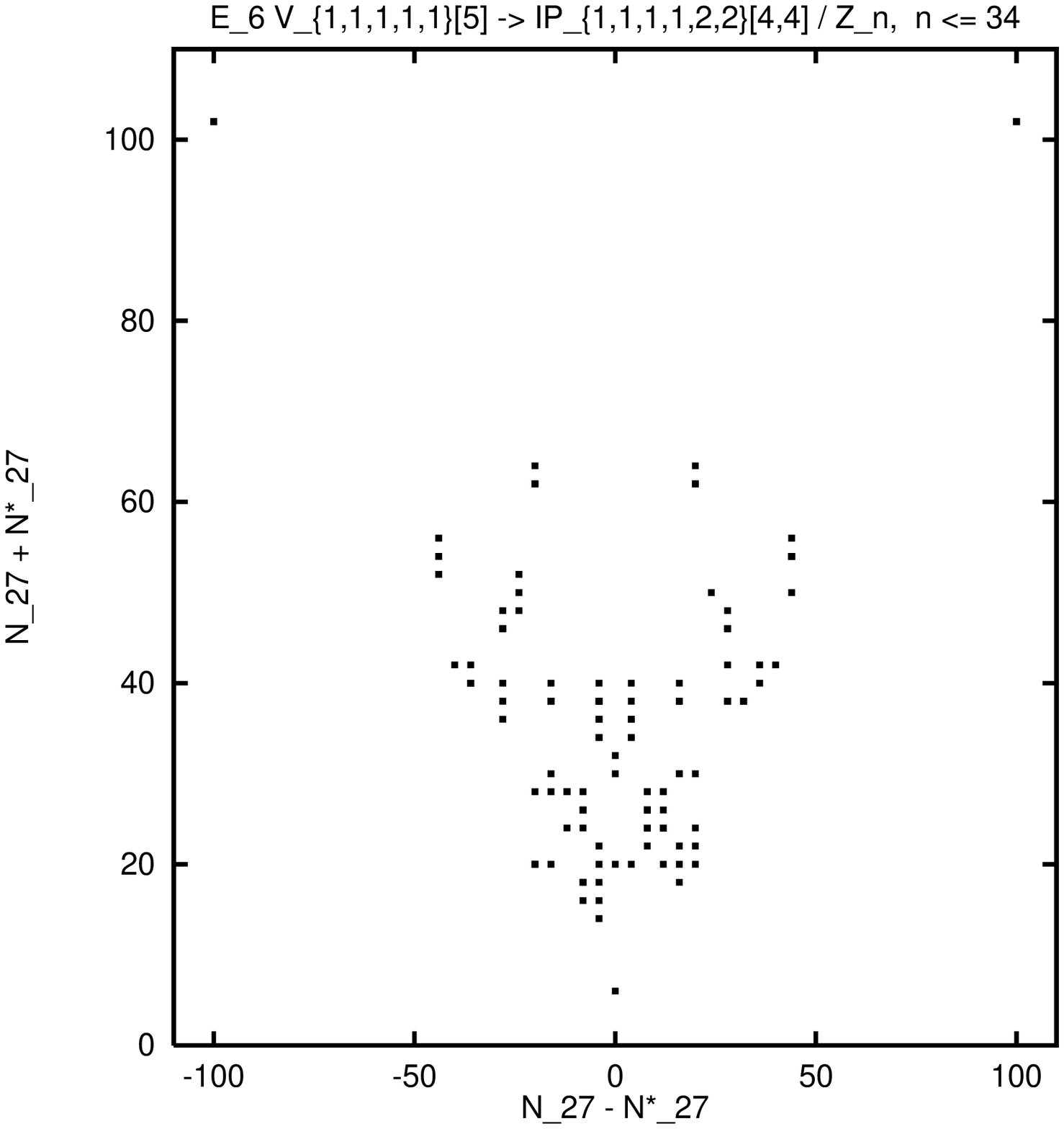}{8.4truecm}

\subsec{Mirror symmetry for general $(0,2)$ models}

Mirror symmetry might be something to be expected for $(0,2)$ descendants
of $(2,2)$ models. Therefore, it is natural to look for orbifolds of a 
general $(0,2)$ model which is not a $(2,2)$ descendant. 
As an example for this, let us take the $(N_{16},N_{\o{16}})=(75,1)$ model
\eqn\modella{ V(1,1,1,1,4;8)\lra \IP_{1,1,2,2,3,3}[6,6] }
with the following choice for the superpotential:
\eqn\superquint{ W=\lambda_1\phi_1^7 +\lambda_2\phi_2^7 
                  +\lambda_3\phi_3^2\phi_5 +\lambda_4\phi_4^2\phi_6
                  +\lambda_5(\phi_3^2 + \phi_4^2)
                  +\lambda_6(\phi_3^3 + \phi_5^2)
                  +\lambda_7(\phi_4^3 + \phi_6^2)\,. }
We did a similar search for this model with symmetry groups $\ZZ_n$,
$2\leq n\leq 40$, and $n=48$. There are some notable differences between
this model and the quintic: The superpotential is much more complicated and
has much less inner symmetry. Therefore, there are much less solutions of
$\ZZ_n$ charges such that $W$ remains invariant and all other consistency
requirements are met. For example, there are no non-trivial solutions for 
$n=9,13,17,18,19,23,25,29,31,34,35,36$, and the total
number of possible solutions in our range is only $5172 + 6208 = 11380$
orbifolds, where the $6208$ models all have $\ZZ_{48}$ symmetry. In total
we only get $192$ different spectra, which in the plot are not differentiated
according to their gauge group.
\pano
The plot (Figure 3) shows much less mirror symmetry than the plot for the
$(0,2)$ quintic, but still a certain amount of it.
Keeping in mind that solutions are harder to find for this example, and that
the naive overall symmetry of $W$ is $\ZZ_{24}$, we might expect a complete
set of solutions only within a huge range $\ZZ_n$, $n\leq 24^k$, with an
unknown power $k\geq 2$.
In the case of the quintic our computation abilities were good enough to get
all orbifolds which could also have been obtained by modding out twice with
basic $\ZZ_5$ symmetries. In this example, however, we are far from such
a degree of completeness. Hence, we might take the appearance of a glance
of symmetry in the plot as a hint that mirror symmetry might be true for
general $(0,2)$ models. This would mean that mirror symmetry is a structure
not just inherited  from $(2,2)$ models, but much deeper and general.

\fig{{\it ~~Orbifolds for an example of a $(0,2)$ model which is {\it not\/} 
a $(2,2)$ descendant.}}{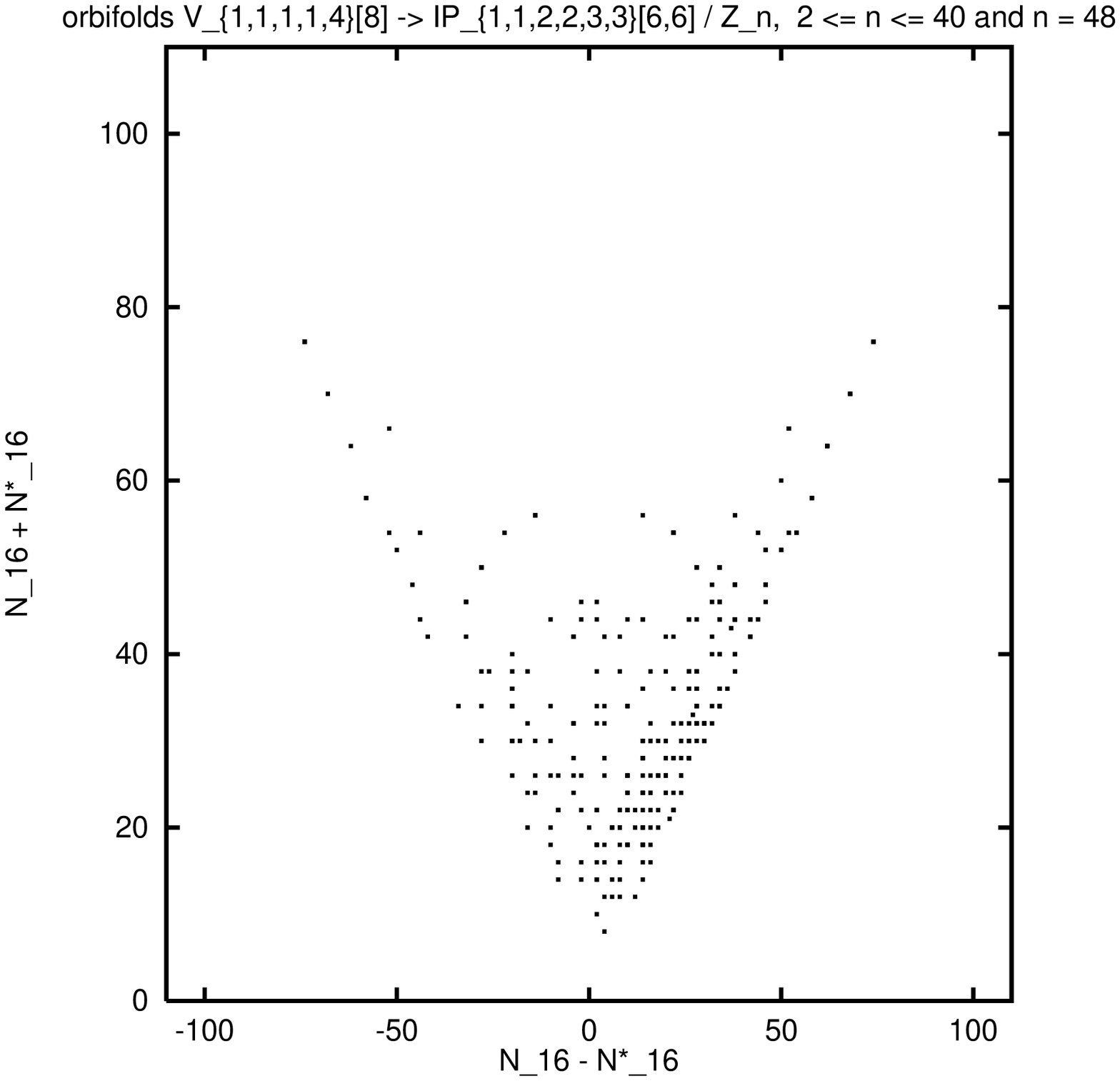}{9.6truecm}

\newsec{Yukawa couplings}

Using left moving $N=2$ supersymmetry, it can be shown that the Yukawa 
coupling $\la 27,27,27\ra$ is independent of the K\"ahler modulus \rdix\rdgb\
and in particular does not receive world sheet instanton corrections.
By choosing a $(2,2)$ model as the internal sector of a type II 
compactification one recognizes this independence to be equivalent to the 
fact that for $N=2$ four dimensional
theories the hyper  and vector multiplets decouple. However, in the general
$(0,2)$ case, there is no left moving supersymmetry and one can not in quite
generality expect a similar property to hold. Since in the $(2,2)$ case the
chiral multiplets in the $\bf{27}$ representation of $E_6$ are related to the 
complex moduli by left supersymmetry, the complex moduli space can also 
be calculated at $\si$-model  tree level. Thus, one gets a complete picture
of the complex and K\"ahler moduli space by looking at the complex moduli
space of the original model and its mirror. Thus, $(2,2)$ mirror symmetry
is not only an abstract  duality but also has far reaching computational
consequences. 
\pano 
The question we are facing in this section is, whether similar applications
of mirror symmetry hold in the $(0,2)$ case.
Due to lack of left supersymmetry the chain of arguments above fails at 
every single step. So we have to be more modest. We consider models
with gauge group $SO(10)$ having a well behaving Landau Ginzburg phase for
$r\ll 0$.  Then, we know that at the Landau Ginzburg point we can define a
chiral ring structure for the Yukawa couplings $\la 10,16,16\ra_{ut}$ in the
untwisted sector \rbsw. This means that the chiral multiplets both in the 
spinor and in the vector representation of $SO(10)$ are given by polynomials
in the zero modes $\phi^i_0$. Clearly, the coupling depends on the (unknown) 
normalizations of the vertex operators and the complex and bundle moduli, 
but nevertheless the chiral ring 
implies strong selection rules for such couplings to be non-zero. Taking into
account that mirror symmetry exchanges generations and anti-generations and 
that all anti-generations occur only in twisted sectors, one expects that at 
least some $\la 10,\o{16},\o{16}\ra_{tw}$  Yukawas do also
have a chiral ring structure.  The selection rules then follow from  
the $\la 10,16,16\ra_{ut}$ couplings of the mirror model. 
A very simple example is the model given in the Calabi-Yau phase by the
bundle
\eqn\sequence{0\to\ V\to\bigoplus_{a=1}^{5}{\cal O}(1)\to{\cal O}(5)\to0\,,}
over the threefold configuration $\IP_{(1,1,1,1,2,2)}[4,4]$.
In the \lg phase, the massless sector   contains  $N_{16}=80$ untwisted 
chiral multiplets which transform in the spinor representation of $SO(10)$. 
These are given by polynomials in the $\phi_i$ of degree five modulo the 
seven constraints of weight four. There are no states transforming in the
conjugate spinor representation, and there are $N_{10}=72_{ut}$ untwisted 
and $N_{10}=2_{tw}$ twisted chiral multiplets which transform in the vector 
representation. The untwisted ones are given by polynomials of degree ten 
modulo the constraints. The mirror of this model can be written as the \lg 
phase of 
\eqn\mirror{ V(51,64,60,80,65;360)\lra \IP_{51,60,80,65,128,128}[256,256] }
with $N_{\o{16}}=0$ and $N_{10}=2_{ut}$ and $N_{10}=72_{tw}$ from the 
untwisted and
twisted sector, respectively. Mirror symmetry then implies that the Yukawas
$\la 10,\o{16},\o{16}\ra_{tw}$ of the mirror model \mirror\ satisfy selection
rules given by the polynomial ring of the original model
\eqn\ring{ {\cal R}={\CC(\phi_i) \over W_j=F_a=0 }\,.}
Since this example is not of a fairly general type and involves plenty of 
massless fields we will consider a different example in more detail, namely
\eqn\modella{ V(1,1,3,5,5;15)\lra \IP_{1,1,6,6,5,5}[12,12]\,. }
The massless spectrum consists of $(N_{16},N_{\o{16}})=(80_{ut},8_{tw})$
generations and anti-generations and $N_{10}=75_{ut} + 7_{tw}$ vectors.
We choose the superpotential to be
\eqn\supot{ W=\lambda_1\phi_1^{14} + \lambda_2\phi_2^{14} 
             + \lambda_3\phi_3\phi_4
             + \lambda_4 \phi_5^2 + \lambda_6 \phi_6^2 + \lambda_6 \phi_3^2 
             + \lambda_7 \phi_4^2\,. }
The mirror is given by taking the $\ZZ_{15}$ quotient acting as
\eqn\chargeb{ (\vec{q}_i;\vec{q}_a)=\left({1\over 15},-{1\over 15},0,0,0,0;
            -{1\over 15},{1\over 15},0,0,0,0,0\right)\,. } 
with  spectrum $(N_{16},N_{\o{16}})=(6_{ut} + 2_{tw},80_{tw})$ and
$N_{10}=5_{ut} + 77_{tw}$. The untwisted sector of the mirror 
contains information about the $\la 10,\o{16},\o{16}\ra_{tw}$ couplings
of the original model. To be more precise, the $N_{16}=6_{ut}$ untwisted 
states are represented by polynomials $A=\phi_1^5\phi_2^5\phi_{5,6}$ and
$B=\phi_1^2\phi_2^2\phi_{3,4}\phi_{5,6}$ of degree fifteen 
and the $N_{10}=5_{ut}$ untwisted vectors
are represented by polynomials $S=\phi_1^{12}\phi_2^{12}\phi_{3,4}$, 
$T=\phi_1^{10}\phi_2^{10}\phi_{5}\phi_{6}$ and 
$U=\phi_1^7\phi_2^7\phi_{3,4}\phi_{5}\phi_6$ of degree thirty. 
The polynomial ring then tells us that couplings do only have a chance
to be non-zero if they are of type $TAA$ or $UAB$ with the indices chosen
appropriately. In order to check this, one would have to calculate the
$\la 10,\o{16},\o{16}\ra_{tw}$ couplings in the original model exactly.
Fortunately, for this model a conformal field theory description is known:
One starts with the $(1,1,3,13,13)$ Gepner model and introduces the 
simple current
\eqn\sc{  J=(0\ 0\ 0)^2 (0\ 5\ 1)(0\ 0\ 0)^2 (1)(0) }
into the partition function. Following the discussion in \rbsw\ 
determining  the massless states and calculating
the Yukawa couplings really shows that there are six anti-generations
and five twisted vectors obeying exactly  the selection rules above.
The exact calculation yields for the couplings
\eqn\exactyu{ TAA={\Gamma\left({1\over 15}\right) 
                   \Gamma^2\left({3\over 5}\right) 
                   \Gamma\left({11\over 15}\right) 
                   \over 
                   \Gamma\left({4\over 15}\right) 
                   \Gamma^2\left({2\over 5}\right) 
                   \Gamma\left({14\over 15}\right)  }\,, \quad
              UAB={\Gamma\left({1\over 15}\right) 
                   \Gamma\left({8\over 15}\right) 
                   \Gamma\left({3\over 5}\right) 
                   \Gamma\left({4\over 5}\right)
                   \over 
                   \Gamma\left({1\over 5}\right) 
                   \Gamma\left({2\over 5}\right)
                   \Gamma\left({7\over 15}\right)
                   \Gamma\left({14\over 15}\right) 
}\,.}
This easy example shows that one can indeed learn one bit of information
{}from $(0,2)$ mirror symmetry. It remains to be seen whether for perhaps a
subclass like all linear $\si$-models stronger statements can be made. 
As was nicely shown in \rsw\ there are unexpected cancellations in the
space-time superpotential so that at least all parameters in a 
linear $\si$-model are indeed good moduli. Similar mechanisms are perhaps at
work to cancel various corrections for the Yukawa couplings.  

Summarizing, we have seen that $(0,2)$ orbifolding is a powerful method to 
get new $(0,2)$ vacua of the heterotic string. Furthermore, we found strong 
indications that descendant $(0,2)$ models of $(2,2)$ models feature $(0,2)$ 
mirror symmetry.
Finally, we argued that $(0,2)$ mirror symmetry can be used to extract
information about couplings of  type $\la 10,\o{16},\o{16}\ra_{tw}$ without
knowing the exact conformal field theory.
\vfill\eject

\listrefs

\bye
\end